\documentclass[aps,twocolumn,amsmath,amssymb,floatfix,showpacs,
superscriptaddress]{revtex4}

\usepackage[dvips]{graphics}

\begin{document}

\title{Magnetic Response of Interacting Electrons in a Fractal Network: 
A Mean Field Approach}

\author{Santanu K. Maiti}

\affiliation{Department of Physics, Narasinha Dutt College, 129
Belilious Road, Howrah-711 101, India}

\author{Arunava Chakrabarti}

\affiliation{Department of Physics, University of Kalyani, Kalyani,
West Bengal-741 235, India.}

\begin{abstract}
The Hubbard model on a Sierpinski gasket fractal is carefully examined 
within a Hartree-Fock mean field approach. We examine  the influence 
of a magnetic flux threading the gasket on its ground state energy, 
persistent current and the Drude weight. Both an isotropic gasket and 
its anisotropic counterpart have been examined. The variance in the 
patterns of the calculated physical quantities are discussed for two 
situations, viz, at half-filling and when the `band' is less than 
half-filled. The phase reversal of the persistent currents and the 
change of the Drude weight as a function of the Hubbard interaction 
are found to exhibit interesting patterns that have so far remained 
unaddressed.  
\end{abstract}


\maketitle

\section{Introduction}
Deterministic fractals have been known to bridge the gap between systems 
possessing perfect periodic order and the completely random ones. The 
spectrum of non interacting electrons on such lattices has been 
exhaustively investigated in the past~\cite{domany,rammal,banavar,ghez,
maritan,gordon1,gordon2,gordon3,schwalm1,andrade1,schwalm2,andrade2,
schwalm3,kappertz,lin,wang1,andrade3,hu,andrade4,macia,korshu,meyer,new}.  
The principal characteristic features of a deterministic fractal may be 
summarized as follows: First, the energy spectrum is a Cantor set, and 
its degenerate~\cite{domany}. Second, the density of states displays a  
variety of singularities and a magnetic field is shown to broaden up 
the spectrum ~\cite{banavar,ghez}, and third, the electronic conductance  
exhibits scaling with a multi-fractal distribution of the 
exponents~\cite{schwalm2}. Apart from these, in certain cases, isolated 
extended eigenstates also appear in deterministic, finitely ramified 
fractal lattices~\cite{wang2,arun1,arun2,arun3}, and extensive numerical 
work has recently proposed a possible existence of even a continuum of 
such extended states~\cite{schwalm4}. 

However, the  typical properties exhibited by the deterministic fractals 
are obtained within the picture of non-interacting spinless Fermions.  
The very fundamental questions such as whether the spectral peculiarities 
exist even in the presence of say, electron-electron interaction, or 
whether the response of a fractal lattice to an externally applied 
magnetic (or electric) field brings out any new features when one looks 
beyond the non-interacting picture, are still to be addressed. The effect 
of electron-electron interaction on the spectral properties are, to our 
mind, is of great importance, particularly because of several experiments 
done on fractal networks that studied the magnetoresistance, the 
superconductor-normal phase boundaries on Sierpinski gasket wire 
networks~\cite{gordon1,gordon2,gordon3,korshu,meyer}. These experiments, 
together with the earlier ones on regular square or honeycomb 
networks~\cite{pannet1,pannet2} to study the flux quantization effects 
pioneered the actual observational studies of spectral properties of 
planar networks and the Aharonov-Bohm effect in systems with or without 
translational invariance. Although in an early paper the problem of 
interacting electrons on a percolating cluster that displays a fractal 
geometry~\cite{nedellec}, has been addressed, to the best of our knowledge, 
no rigorous effort has been made so far to unravel the effect of an 
interplay of electron-electron interaction and an external magnetic field 
on deterministic networks such as a Sierpinski gasket (SPG), even at a 
mean field level.
 
This inspires us to undertake a detailed study of the ground state energy 
and the magnetic response of a Sierpinski gasket (SPG) fractal~\cite{domany,
rammal,banavar} that stands out to be a classic example of such lattices, 
and has been the subject of the experiments cited above. We examine the 
persistent current~\cite{chung1,chung2} in such a fractal in the presence 
of on-site Hubbard interaction within an unrestricted Hartree-Fock mean 
field scheme. Persistent current in normal metal loops~\cite{chung1,chung2,
georges,santanu} is an important effect in mesoscopic dimensions. Here, 
an SPG network offers a unique opportunity to study the persistent current 
in a self-similar distribution of loops, and with correlated electrons it 
is likely to give rise to new observations. This is a major motivation of 
the present work.

Apart from this, the magnetoconductance (Drude weight) has also been 
calculated and the variation of the response of the lattice to the 
external magnetic field has been carefully studied as the fractal 
grows in size. To the best of our knowledge the interplay of a fractal 
geometry and electron-electron correlation in the form of persistent 
currents and the Drude weight has not been studied before. With the 
metallic SPG networks already synthesized, the present study may motivate 
experiments for a direct observation of the effects presented here. 
In particular, based on the success of the lithographic techniques it 
may not be too wild an idea to suggest an SPG kind of fractal network 
built by carbon nanotubes that are connected at the vertices. 

As mentioned before, we examine both the isotropic and the anisotropic 
SPG fractal networks. The anisotropy is introduced only in the values of 
the nearest-neighbor hopping integrals. The response of the lattice is 
found to differ grossly for an anisotropic system compared to the 
isotropic one. This is of course, dependent on the relative values of 
the parameters in the Hamiltonian, through which the anisotropy enters 
the system. For example, the anisotropic SPG fractal is found to be more 
conducting than the isotropic one in the sense that, the lattice remains 
conducting over a wide range of values of the Hubbard interaction. The 
magnitude of the conductivity however, is sensitive to the strength of 
the hopping parameters. This fact has also been reported recently for 
non-interacting electrons~\cite{jana}.

In what follows, we present the results. In section II, the model 
Hamiltonian is presented. Section III briefly describes the mean field 
approach, while the results and the discussion are included in section 
IV. In section V we draw our conclusions.

\section{The Model}

\begin{figure}[ht]
{\centering \resizebox*{5cm}{5cm}{\includegraphics{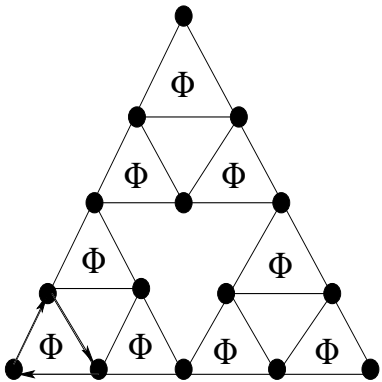}}\par}
\caption{A $3$-rd generation Sierpinski gasket in which each elementary 
plaquette is penetrated by a magnetic flux $\phi$. The filled black 
circles correspond to the positions of the atomic sites.}  
\label{gasket}
\end{figure}

We start by referring to Fig.~\ref{gasket} where a $3$-rd generation 
SPG in which each elementary plaquette is threaded by a magnetic flux 
$\phi$ (measured in unit of the elementary flux quantum $\phi_0=ch/e$) 
is shown. The filled black circles correspond to the positions of the 
atomic sites in the SPG. To describe the system we use a tight-binding 
framework. In a Wannier basis the Hamiltonian reads, 
\begin{eqnarray}
H_{\mbox{SPG}} & = &\sum_{i,\sigma}\epsilon_{i\sigma} c_{i\sigma}^{\dagger} 
c_{i\sigma} + \sum_{\langle ij \rangle,\sigma} t \left[e^{i\theta} 
c_{i\sigma}^{\dagger} c_{j\sigma} + h.c. \right] \nonumber \\
&  & +\sum_i U c_{i\uparrow}^{\dagger}c_{i\uparrow} c_{i\downarrow}^{\dagger}
c_{i\downarrow}
\label{equ1}
\end{eqnarray}
where, $\epsilon_{i\sigma}$ is the on-site energy of an electron at 
the site $i$ of spin $\sigma$ ($\uparrow,\downarrow$) and $t$ is the 
nearest-neighbor hopping strength. In the case of an anisotropic SPG, 
the anisotropy is introduced only in the nearest-neighbor hopping 
integral $t$ which takes on values $t_x$ and $t_y$ for hopping along 
the {\it horizontal} and the {\it angular} bonds, respectively. Due to 
the presence of magnetic flux $\phi$, a phase factor $\theta=2\pi\phi/3$ 
appears in the Hamiltonian when an electron hops from one site to another 
site, and accordingly, a negative sign comes when the electron hops in 
the reverse direction. As the magnetic filed associated with the flux 
$\phi$ does not penetrate any part of the circumference of the elementary 
triangle, we ignore the Zeeman term in the above tight-binding 
Hamiltonian (Eq.~\ref{equ1}). $c_{i\sigma}^{\dagger}$ and $c_{i\sigma}$ 
are the creation and annihilation operators, respectively, of an electron 
at the site $i$ with spin $\sigma$. $U$ is the strength of on-site Coulomb 
interaction. 

\section{The mean field approach}

\subsection{Decoupling of the interacting Hamiltonian}

To determine the energy eigenvalues of the interacting model of the SPG 
described by the tight-binding Hamiltonian given in Eq.~\ref{equ1}, first 
we decouple the interacting Hamiltonian using the generalized Hartree-Fock 
approach~\cite{kato,kam}. The full Hamiltonian is completely decoupled 
into two parts. One is associated with the up-spin electrons, while the 
other is with the down-spin electrons. The on-site potentials get modified 
appropriately, and are given by,
\begin{equation}
\epsilon_{i\uparrow}^{\prime}=\epsilon_{i\uparrow} + U \langle 
n_{i\downarrow} \rangle
\label{equ2}
\end{equation}
\begin{equation}
\epsilon_{i\downarrow}^{\prime}=\epsilon_{i\downarrow} + U \langle 
n_{i\uparrow} \rangle
\label{equ3}
\end{equation}
where, $n_{i\sigma}=c_{i\sigma}^{\dagger} c_{i\sigma}$ is the number
operator. With these site energies, the full Hamiltonian (Eq.~\ref{equ1})
can be written in the decoupled form (in the mean field approximation) as,
\begin{eqnarray}
H_{\mbox{mean field}} &=&\sum_i \epsilon_{i\uparrow}^{\prime} n_{i\uparrow} + 
\sum_{\langle ij \rangle} t \left[e^{i\theta} c_{i\uparrow}^{\dagger} 
c_{j\uparrow} + e^{-i\theta} c_{j\uparrow}^{\dagger} 
c_{i\uparrow}\right] \nonumber \\
& + & \sum_i \epsilon_{i\downarrow}^{\prime} n_{i\downarrow} + \sum_{\langle 
ij \rangle} t \left[e^{i\theta} c_{i\downarrow}^{\dagger} c_{j\downarrow}
+ e^{-i\theta} c_{j\downarrow}^{\dagger} c_{i\downarrow}\right] \nonumber \\
& - & \sum_i U \langle n_{i\uparrow} \rangle \langle n_{i\downarrow} 
\rangle \nonumber \\
&=& H_{\uparrow}+H_{\downarrow}-\sum_i U \langle n_{i\uparrow} 
\rangle \langle n_{i\downarrow} \rangle
\label{equ4} 
\end{eqnarray}
where, $H_{\uparrow}$ and $H_{\downarrow}$ correspond to the effective
tight-binding Hamiltonians for the up and down spin electrons, respectively.
The last term is a constant term which provides a shift in the 
total energy. 

\subsection{Self consistent procedure}

With these decoupled Hamiltonians ($H_{\uparrow}$ and $H_{\downarrow}$) 
of up and down spin electrons, now we start our self consistent procedure 
considering initial guess values of $\langle n_{i\uparrow} \rangle$ and 
$\langle n_{i\downarrow} \rangle$. For these initial set of values of
$\langle n_{i\uparrow} \rangle$ and $\langle n_{i\downarrow} \rangle$, 
we numerically diagonalize the up and down spin Hamiltonians. Then we 
calculate a new set of values of $\langle n_{i\uparrow} \rangle$ and 
$\langle n_{i\downarrow} \rangle$. These steps are repeated until a self
consistent solution is achieved.

\subsection{The ground state energy}

After achieving the self consistent solution, the ground state energy 
$E_0$ for a particular filling at absolute zero temperature ($T=0$K)
can be determined by taking the sum of individual states up to the Fermi 
energy ($E_F$) for both the up and down spins. The final expression of 
the ground state energy is written,
\begin{equation}
E_0=\sum_n E_{n\uparrow} + \sum_n E_{n\downarrow}- \sum_i U \langle
n_{i\uparrow} \rangle \langle n_{i\downarrow} \rangle
\label{equ5} 
\end{equation} 
where, the index $n$ runs over the states up to the Fermi level. 
$E_{n\uparrow}$ ($E_{n\downarrow}$) is the single particle energy 
eigenvalue for $n$-th eigenstate obtained by diagonalizing the Hamiltonian 
$H_{\uparrow}$ ($H_{\downarrow}$).

\subsection{Calculation of persistent current} 

At absolute zero temperature, total persistent current of the system
is obtained from the expression~\cite{chung1,chung2}
\begin{equation}
I(\phi)=-c\frac{\partial E_0(\phi)}{\partial \phi}
\label{equ6}
\end{equation}
where, $E_0(\phi)$ is the ground state energy for a particular filling.

\subsection{Calculation of Drude weight} 

The conductance can be obtained by calculating the Drude weight $D$ as 
originally noted by Kohn~\cite{kohn}. The Drude weight for the SPG is 
obtained through the relation,
\begin{equation}
D=\left . \frac{N}{4\pi^2} \left(\frac{\partial{^2E_0(\phi)}}
{\partial{\phi}^{2}}\right) \right|_{\phi \rightarrow 0}
\label{equ7}
\end{equation}
where, $N$ gives total number of atomic sites in the gasket. Kohn 
has shown that for an insulating system $D$ decays exponentially
to zero, while it becomes finite for a conducting system.

In the present work we inspect all the essential features of magnetic
response of an SPG network at absolute zero temperature and use the 
units where $c=h=e=1$. Throughout our numerical work we set 
$\epsilon_{i\uparrow}=\epsilon_{i\downarrow}=0$ for all $i$ and choose 
the nearest-neighbor hopping strength $t=-1$. In the anisotropic case 
we select $t_x = -1$ and $t_y = -2$ throughout. Energy scale is measured 
in unit of $t$. Results are obtained both for an isotropic gasket and 
its anisotropic counterpart.

\section{Numerical results and discussion}

\begin{figure}[ht]
{\centering \resizebox*{7.75cm}{7cm}
{\includegraphics{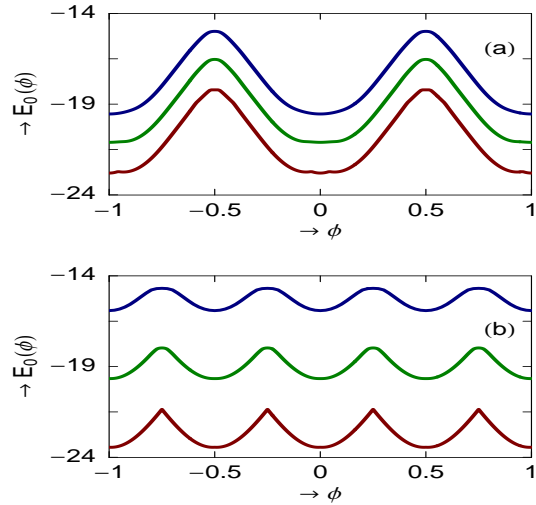}}\par}
\caption{(Color online). Ground state energy levels as a function of 
flux $\phi$ for a $3$-rd generation isotropic ($t_x=t_y=-1$) Sierpinski 
gasket ($N=15$). The red, green and blue curves correspond to $U=0$, 
$1$ and $2$, respectively. (a) $N_e=10$ and (b) $N_e=15$.}
\label{fractalenergy}
\end{figure}

In Fig.~\ref{fractalenergy} we present the variation of the ground state 
energy of a $3$-rd generation isotropic SPG containing $15$ atomic sites
as a function of the magnetic flux through each elementary triangle. Two 
cases, viz, when the `band' is less that half-filled, and half-filled, 
\begin{figure}[ht]
{\centering \resizebox*{7.75cm}{7cm}
{\includegraphics{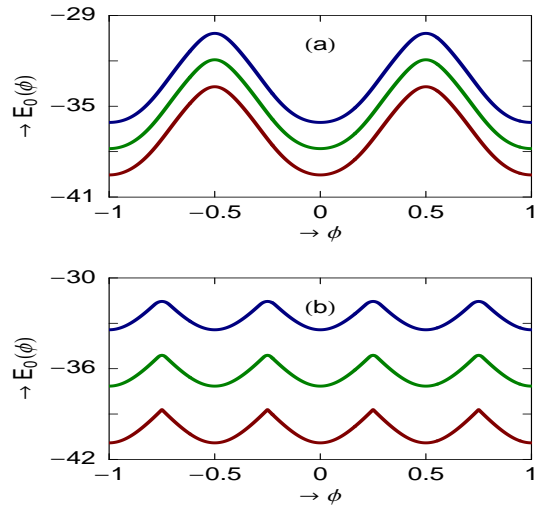}}\par}
\caption{(Color online). Ground state energy levels as a function of 
flux $\phi$ for a $3$-rd generation anisotropic ($t_x=-1$ and $t_y=-2$) 
Sierpinski gasket ($N=15$). The red, green and blue curves correspond 
to $U=0$, $1$ and $2$, respectively. (a) $N_e=10$ and (b) $N_e=15$.}
\label{fractalenergy1}
\end{figure}
are presented as the on-site Coulomb repulsion $U$ is varied. The ground 
state energy exhibits a periodicity equal to one flux quantum in all the 
non-half-filled cases, while the period changes to half flux quantum at 
precisely half-filling. With increasing $U$, the ground state energy 
increases in both these cases. In the half-filled case, each site is 
occupied by at least one electron, and the placing of a second electron 
will increase the energy of the system (the effect of $U$). This is 
\begin{figure}[ht]
{\centering \resizebox*{7.75cm}{7cm}
{\includegraphics{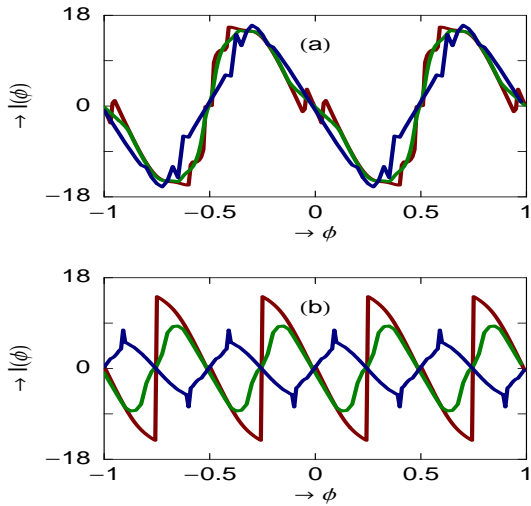}}\par}
\caption{(Color online). Persistent current as a function of flux $\phi$ 
for a $3$-rd generation isotropic ($t_x=t_y=-1$) Sierpinski gasket ($N=15$). 
The red, green and blue curves correspond to $U=0$, $2$ and $4$, 
respectively. (a) $N_e=10$ and (b) $N_e=15$.}
\label{fractalcurr}
\end{figure}
\begin{figure}[ht]
{\centering \resizebox*{7.75cm}{7cm}
{\includegraphics{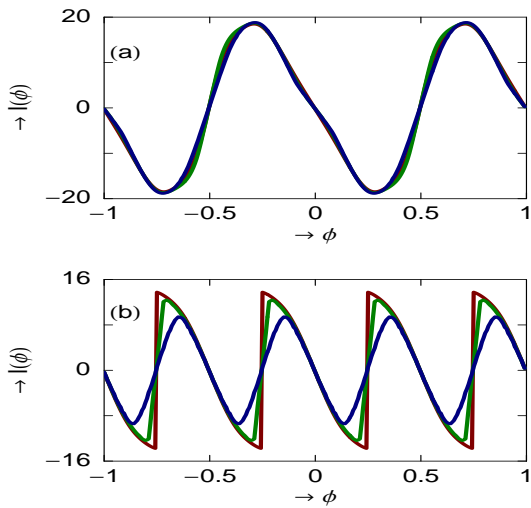}}\par}
\caption{(Color online). Persistent current as a function of flux $\phi$ 
for a $3$-rd generation anisotropic ($t_x=-1$ and $t_y=-2$) Sierpinski 
gasket ($N=15$). The red, green and blue curves correspond to $U=0$, $2$ 
and $4$, respectively. (a) $N_e=10$ and (b) $N_e=15$.}
\label{fractalcurr1}
\end{figure}
reflected in Fig.~\ref{fractalenergy}(b). Also the values of the ground 
state energy in the half-filled case turns out to be well separated from 
each other for $U = 0$, $1$ and $2$ compared to the non-half-filled case 
in (a). This feature remains true irrespective of the size of the system.

As anisotropy is introduced, the overall features remain unaltered, 
including the periodicities. However, as is evident from 
Fig.~\ref{fractalenergy1}, the anisotropy lowers the ground state energy 
of an SPG, both in the non-half-filled and the half-filled cases. This 
will be reflected in the conductance, as will be shown later.

The variation of the persistent current against the magnetic flux is 
shown separately for the isotropic (Fig.~\ref{fractalcurr}) and the 
anisotropic (Fig.~\ref{fractalcurr1}) SPG for different values of the 
Hubbard interaction $U$. Two typical results, when $N_e = 10$ (less 
than half-filled case) and $N_e = 15$ (half-filling), are presented for 
a third generation SPG with $N = 15$ sites. In Fig.~\ref{fractalcurr}(a) 
and in Fig.~\ref{fractalcurr1}(a) results for the `less than half-filled' 
case are presented. In Fig.~\ref{fractalcurr}(a) the $I(\phi)$-$\phi$ 
curves exhibit multiple kinks which follows from the numerous 
\begin{figure}[ht]
{\centering \resizebox*{7.75cm}{8cm}
{\includegraphics{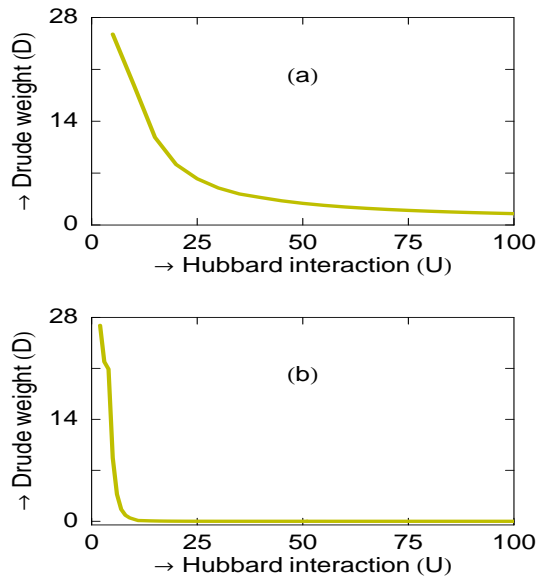}}\par}
\caption{(Color online). Drude weight as a function of Hubbard interaction
strength $U$ for a $3$-rd generation isotropic ($t_x=t_y=-1$) Sierpinski 
gasket ($N=15$). (a) Non-half-filled case ($N_e=10$). (b) Half-filled case 
($N_e=15$).}
\label{fractaldrude}
\end{figure}
band-crossings that are typical of such hierarchical 
networks~\cite{banavar,jana}. Such crossings become less in number, 
and global gaps open up in the spectrum, clustering the spectrum into 
sub-band structures in the case of an anisotropic SPG, as has recently 
been reported in the literature even in the case of non-interacting 
electrons~\cite{jana}. Kinks are now expected to smooth out. That it 
happens, is evident from the anisotropic case, as depicted in 
Fig.~\ref{fractalcurr1}(a). So, anisotropy turns out to be the 
predominant factor in reducing the band-crossings here.

On the other hand, in the half-filled case, the isotropic version of the 
SPG display (Fig.~\ref{fractalcurr}(b)) non-trivial characteristics 
compared to its anisotropic counterpart (Fig.~\ref{fractalcurr1}(b)). 
In the former case the increasing value of $U$ is seen to result into 
a complete reversal of the phase of the persistent current, converting 
a diamagnetic response to a paramagnetic one. This however is not seen 
to happen (in the half-filled case) in an anisotropic gasket 
(Fig.~\ref{fractalcurr1}(b)).

\begin{figure}[ht]
{\centering \resizebox*{7.75cm}{8cm}
{\includegraphics{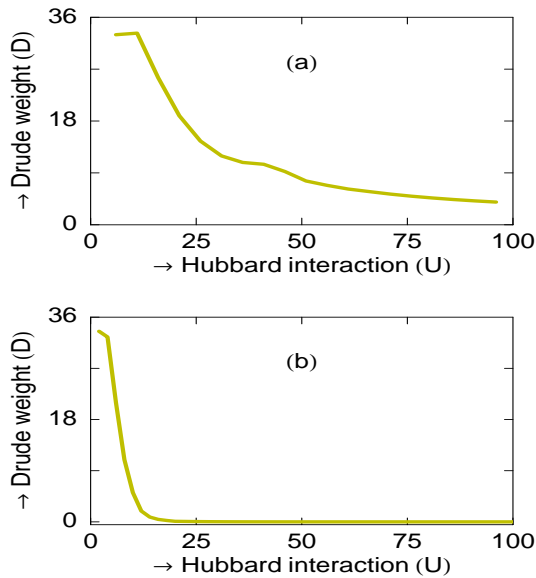}}\par}
\caption{(Color online). Drude weight as a function of Hubbard interaction
strength $U$ for a $3$-rd generation anisotropic ($t_x=-1$ and $t_y=-2$) 
Sierpinski gasket ($N=15$). (a) Non-half-filled case ($N_e=10$). 
(b) Half-filled case ($N_e=15$).}
\label{fractaldrude1}
\end{figure}

We now present the results of the calculation of Drude weight $D$ both 
in the cases of an isotropic and an anisotropic SPG, and observe its 
variation as $U$ increases. Results are presented in 
Fig.~\ref{fractaldrude} and Fig.~\ref{fractaldrude1}, respectively, 
for a $3$-rd generation gasket. It is apparent that, the anisotropic 
gasket turns out to be more conducting than its isotropic counterpart 
in the sense that, in the anisotropic case the Drude weight displays 
finite values over a wider range of $U$. The magnitude of $D$ at any 
specific $U$ of course, depends on the numerical values of the hopping 
strength. Interestingly, this fact is also  
observed~\cite{jana} for non-interacting electrons on an SPG. In the 
half-filled band case, the Drude weight exhibits a much sharper drop 
in its value compared to the non-half-filled situation. It is true 
for both the isotropic as well as the anisotropic case. The reason can 
easily be traced back again to the fact that at half-filling, every 
site of the SPG network has one electron occupying it already. So, 
conduction becomes difficult as one needs more energy when an electron 
tries to leave its own site and occupy a neighboring site. At less than 
half-filling there are `empty' lattice points and conduction becomes 
easier. However, we find that in the anisotropic case, we have to make 
the on-site Hubbard interaction much stronger compared to the isotropic 
case to lower the value of the conductance close to zero. 

Before we end this section, it is pertinent to raise the question as to 
whether the features discussed above really represent the characteristics 
of a fractal. To get a definite answer to this, we have extended our 
analysis to higher generation SPG networks, both in the isotropic and the 
anisotropic limits. In each case, the overall features of the ground state 
energy, the persistent current or the Drude weight turn out to be the same 
as in the cases of lower generations. The effect of a variation of the 
Hubbard interaction essentially plays the same role. The difference in 
the numerical values of the quantities are of course, obvious. To clarify, 
we provide the results of our calculation on a fourth generation SPG network 
comprising of $42$ sites in the anisotropic limit, and in the half-filled 
\begin{figure}[ht]
{\centering \resizebox*{7.75cm}{10cm}
{\includegraphics{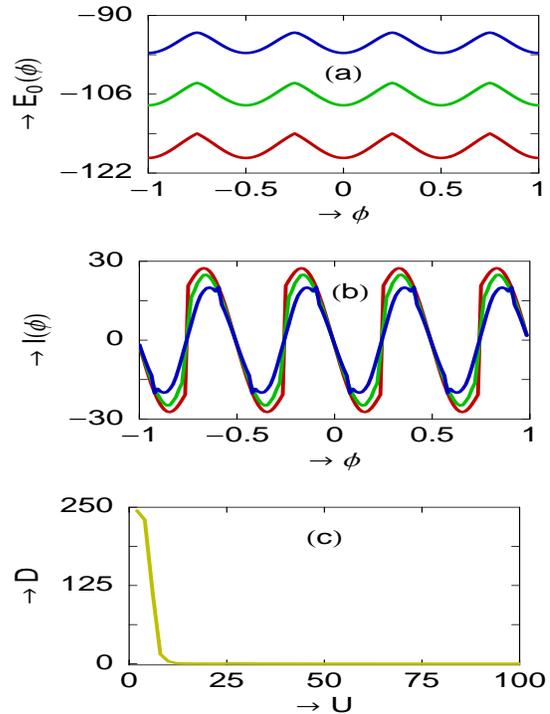}}\par}
\caption{(Color online). Magnetic response for a $4$-th generation 
anisotropic ($t_x=-1$ and $t_y=-2$) Sierpinski gasket ($N=42$) in
Half-filled case ($N_e=42$). (a) Energy-flux characteristics where 
the red, green and blue curves correspond to $U=0$, $1$ and $2$,
respectively. (b) Current-flux characteristics where the red, green 
and blue curves correspond to $U=0$, $2$ and $4$, respectively. 
(c) Drude weight as a function of Hubbard interaction strength.}
\label{fourthgen}
\end{figure}
band case. This is in Fig.~\ref{fourthgen}. The ground state energy in 
this case, as in the previous generations, exhibits the same qualitative 
variation against the magnetic flux, and it is the derivative of the ground 
state energy that generates the current. So, a qualitative similarity 
between the curves at various generations is not unexpected. A direct 
comparison with Fig.~\ref{fractalcurr1} reveals that, the persistent current 
for $U = 0$ in the present case is a bit rounded off at the peak compared 
to the sharp discontinuity exhibited in the corresponding case in the third 
generation fractal. This is not un-natural, as the current depends on the 
band crossings exhibited by the eigenvalue spectrum of the finite generation 
fractals, and the nature of band crossings will change in every generation. 
But, the important point to note is that, the periodicity of the persistent 
current is not affected, and the gradual phase shift shown by the $I(\phi)$ 
curves in every generation, as the Hubbard interaction is increased, is 
consistent. The observations remain the same when we go beyond the fourth 
generation. This attempts us to believe that the features are likely to 
persist for SPG networks of arbitrarily large finite generations.   

\section{Closing Remarks}

In conclusion, we have performed a thorough mean field analysis of the 
response of a Sierpinski gasket fractal to an external magnetic field.
We have examined both the isotropic and the anisotropic limits of the 
system, where the anisotropy is introduced only in the values of the 
nearest-neighbor hopping integrals along two directions. Within the 
framework of the unrestricted Hartree-Fock theory we decouple the Hubbard 
Hamiltonian and obtain the ground state energy, the persistent current 
and the Drude weight. The persistent current exhibits non trivial patterns 
in each case, and even reveals a change in response, from diamagnetic to 
paramagnetic in the isotropic case as a function of the interaction $U$. 
So, the Hubbard interaction is seen to play its part in the magnetic 
response. The band crossing is diminished by the anisotropy. The network 
remains diamagnetic in the isotropic case, as far as we have examined. 
The conductance is obtained through the Drude weight and, depending on 
the values of the nearest-neighbor hopping integrals, the anisotropic 
gasket may remain conducting than its isotropic counterpart
for a wider range of the Hubbard correlation.
\vskip 0.3in
\noindent
{\bf\small ACKNOWLEDGMENTS}
\vskip 0.2in
\noindent
First author thanks Prof. S. N. Karmakar and Prof. Shreekantha Sil for 
illuminating comments and suggestions during the calculations.

\end{document}